\documentclass[twocolumn,showpacs,aps,prl]{revtex4}

\usepackage{graphicx}
\usepackage{amsfonts, amsmath, amstext, amssymb, amsfonts, amsxtra}
\usepackage{braket}
\usepackage{bbm}
\usepackage{bbold}
\usepackage[protrusion=true,expansion=true]{microtype}
\usepackage{wasysym}
\usepackage{cleveref}
\usepackage{color}
\usepackage{ulem}

\newcommand{\im}{{i}}         % imaginary unit
\newcommand{\e}{{e}}          % Euler's number
\newcommand{\id}{\mathbb{1}}  % identity
\begin{document}

\title{Localization in open quantum systems}
\author{I.~Yusipov$^1$, T.~Laptyeva$^2$, S.~Denisov$^{3,4}$, and M.~Ivanchenko$^{4}$}
\affiliation{$^{1}$Institute of Supercomputing Technologies, Lobachevsky University, Gagarina Av.\ 23, Nizhny Novgorod, 603950, Russia \\
$^{2}$Department of Control Theory and Systems Dynamics, Lobachevsky University, Gagarina Av.\ 23, Nizhny Novgorod, 603950, Russia\\
$^{3}$Institute of Physics, University of Augsburg, Universit\"{a}tsstra{\ss}e 1, 86159 Augsburg, Germany \\
$^{4}$Department of Applied Mathematics, Lobachevsky State University of Nizhny Novgorod, Gagarina Av.\ 23, Nizhny Novgorod, 603950, Russia}


\begin{abstract}
In an isolated single-particle quantum system a spatial disorder can induce Anderson localization. 
Being a result of interference, this  phenomenon is expected to be fragile in the face of  dissipation.
Here we show that a proper dissipation can drive a disordered system into a steady state 
with tunable localization properties. 
This can be achieved  with a set of identical dissipative operators, each one acting non-trivially on a pair of 
sites. Operators are parametrized by a uniform phase, which controls selection of Anderson modes  contributing to the state. 
On the microscopic level, quantum trajectories of a system in the asymptotic regime exhibit intermittent dynamics consisting of long-time sticking 
events near selected modes interrupted by inter-mode jumps.  
%Prospects of experimental realization of dissipative localized states are related to polariton condensate 
%lattices and cavity-QED arrays.
\end{abstract}

\maketitle

%
%%%%%%% introduction
Localization by disorder is a fifty-year old phenomenon which is still posing new puzzles and yielding 
new surprises \cite{Kramer1993,Evers2008,fifty}. 
Anderson localization (AL) of non-interacting  particles and waves in quantum systems
%or  waves has been  observed in experiments with electromagnetic, acoustic, and matter waves \cite{Segev2013,Hu2008,Billy2008,Roati2008,Kondov2011,Jen2012}.
%AL in quantum systems 
is well-understood now in the coherent Hamiltonian limit \cite{fifty2,Segev2013,Billy2008,Roati2008,Kondov2011,Jen2012}; however, 
it is less explored in the situation when the systems are \textit{open}, i.e., they interact  with their environments \cite{book}.

An asymptotic localization in open disordered quantum systems might sound like an oxymoron.  
Dissipative effects can, in principle, play a constructive role in bringing  quantum systems into specific  states \cite{DiehlZoller2008, KrausZoller, wolf2009} and
stabilizing them in metastable states \cite{RefB1,RefB2,RefB3}; they  can also be used to inhibit loses and induce coherence in Bose-Einstein condensates \cite{RefA8,RefA10,RefA11,RefA12}.
Yet the phenomenon of AL relies on a fine long-range interference \cite{anderson}, 
and it is intuitively expected that dissipation will blur the latter and thus  eventually
destroy the former. 
%Indeed, it has been demonstrated that scattering and spectral  
%properties of the quantum systems,  exhibiting localization in the coherent limit, are 
%deteriorated under the action of dissipation  \cite{Fyodorov,Huse}.
First evidences that Anderson localization can tolerate  dissipation and survive in the asymptotic limit 
have been  obtained for semi-classical and classical systems. Namely, semi-classical dissipative localization 
is exemplified by a random laser operating in the Anderson regime, where localization reduces the 
spatial overlap and suppresses competition between lasing modes  and thus improving stability  of the laser \cite{Stano2013, LiuJ.2014}. 
Classical active disordered lattices were shown to be able to support so-called `Anderson attractors' \cite{ivanchenko2015a,ivanchenko2015b}.
A recent study of AL in an open quantum system, however, reports eventual destruction of the localization (although on different parameter-dependent time scales) \cite{les0}.

In this Letter we show that a one-dimensional quantum system with a  Hamiltonian exhibiting  Anderson 
localization can be driven into a steady state which bears localization properties.
Such an asymptotic state can be engineered with a set of local dissipative operators, the corresponding mechanism is based on the robust spatial phase-structure of Anderson modes.
This is in the spirit of the `dissipative engineering' \cite{DiehlZoller2008, KrausZoller, wolf2009}; note, however,
that our aim is not to create a pure state but a state (in fact, highly mixed) with desired  localization properties.

We consider the evolution of an open $N$-dimensional quantum system governed by the Lindblad  master equation 
\cite{book,alicki},
\begin{equation}
\label{eq:1}
\dot{\varrho} = \mathcal{L}(\varrho) = -\im [H,\varrho] + \mathcal{D}(\varrho).
\end{equation}
The first term on the r.h.s.\ captures the unitary evolution of the system
governed by  Hamiltonian $H$. The  dissipative part of the Lindblad generator $\mathcal{L}$,
\begin{equation}
\label{dissipator}
\mathcal{D}(\varrho) = \sum_{k=1}^{S} \gamma_{k}(t) \left[V_k\varrho V^\dagger_k - \frac{1}{2}\{V^\dagger_kV_k,\varrho\}\right]
\end{equation}
is built from the set of $S$ operators, $\{V_k\}_{1,...,S}$, which capture action of the environment on the system. 
Under some conditions \cite{alicki}, the  propagator $\mathcal{P}_t
= \e^{\mathcal{L}t}$ is relaxing towards a unique density matrix
$\varrho_{\infty} = \lim_{t \rightarrow \infty} \mathcal{P}_t \varrho_0$ for all $\varrho_0$.
This is a kernel of the  Lindblad generator
in Eq.~(\ref{eq:1}), $\mathcal{L}(\varrho_{\infty}) = 0$.

The asymptotic density matrix $\varrho_{\infty}$ is out-shaped by the joint action of the Hamiltonian and dissipative operators.
If all dissipative operators are Hermitian, $V_k \equiv V^\dagger_k$, this matrix 
is universal and trivial; namely, it is the normalized identity $\varrho_{\infty}  = \id/N$. This is the case considered recently in the context of many-body localization
\cite{fish,les,les2}. Such dissipators do not differentiate between  systems with localization and without,
they are `grinders'. On the other side, formally there is infinitely many choices of non-Hermitian operators $V_k$
which guarantee the asymptotic state in the form  $\varrho_{\infty} = |\phi_i\rangle\langle \phi_i|$, where
$|\phi_i\rangle$ is the $i$-th eigenstate of the Hamiltonian $H$. 
For this $|\phi_i\rangle$ has to be a dark state of all dissipative operators, $V_k |\phi_i\rangle = 0$ \cite{DiehlZoller2008, KrausZoller}.
In practice, however, this will require a priori knowledge of state $|\phi_i\rangle$ and synthesis of  very peculiar 
dissipative operator(s). This can be unfeasible. A realistic and not very specific  choice of operators is more attractive.
Local dissipators, i.e., those which act non-trivially on a finite number of neighboring particles,
are natural in many contexts.
%, each one acting non-trivially on a finite number of neighboring particles (just two in the simplest case)
Such operators is a popular choice in the recent works on  open quantum systems
\cite{DiehlZoller2008, KrausZoller, Bardyn,barr,kienz,ketz}. This is also our choice here.

\begin{figure}
\includegraphics[width=\columnwidth]{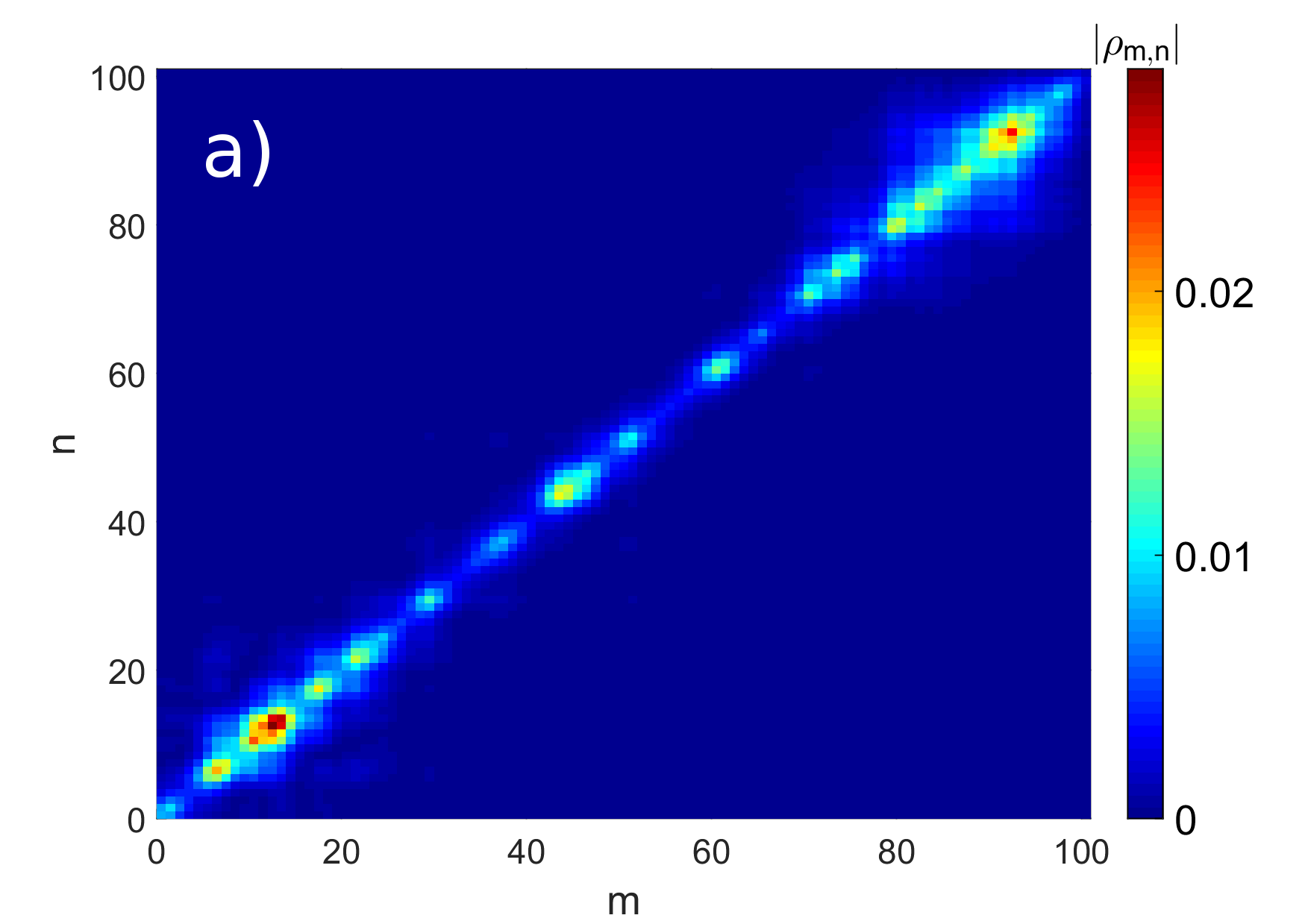}
\includegraphics[width=\columnwidth]{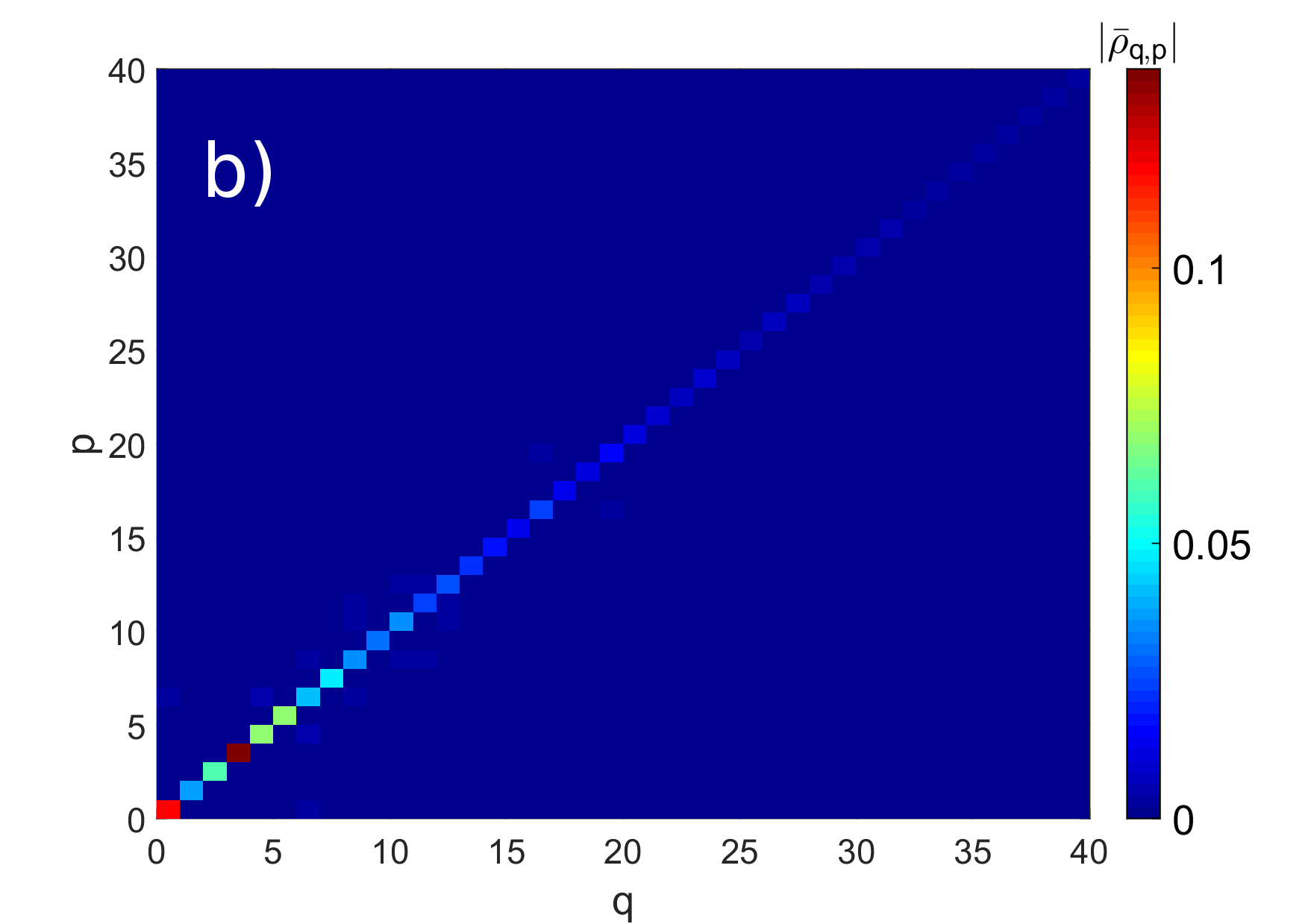}
\caption{Absolute values of the elements of the 
asymptotic density matrix   $\varrho_{\infty}$ expressed in (a) the original basis  and (b) the basis of the Anderson modes (a relevant part is shown) for a particular 
disorder realization of the model system, Eqs.~(\ref{eq:1}-\ref{eq:2}), and
in-phase next-neighbor dissipators, $\alpha  = 0$ and $l=1$ in Eq.~(\ref{eq:4}). Other parameters are $W=1$, $N=100$, and $\gamma =0.1$.} \label{fig:1}
\end{figure}

\begin{figure}
\includegraphics[angle=0,width=\columnwidth]{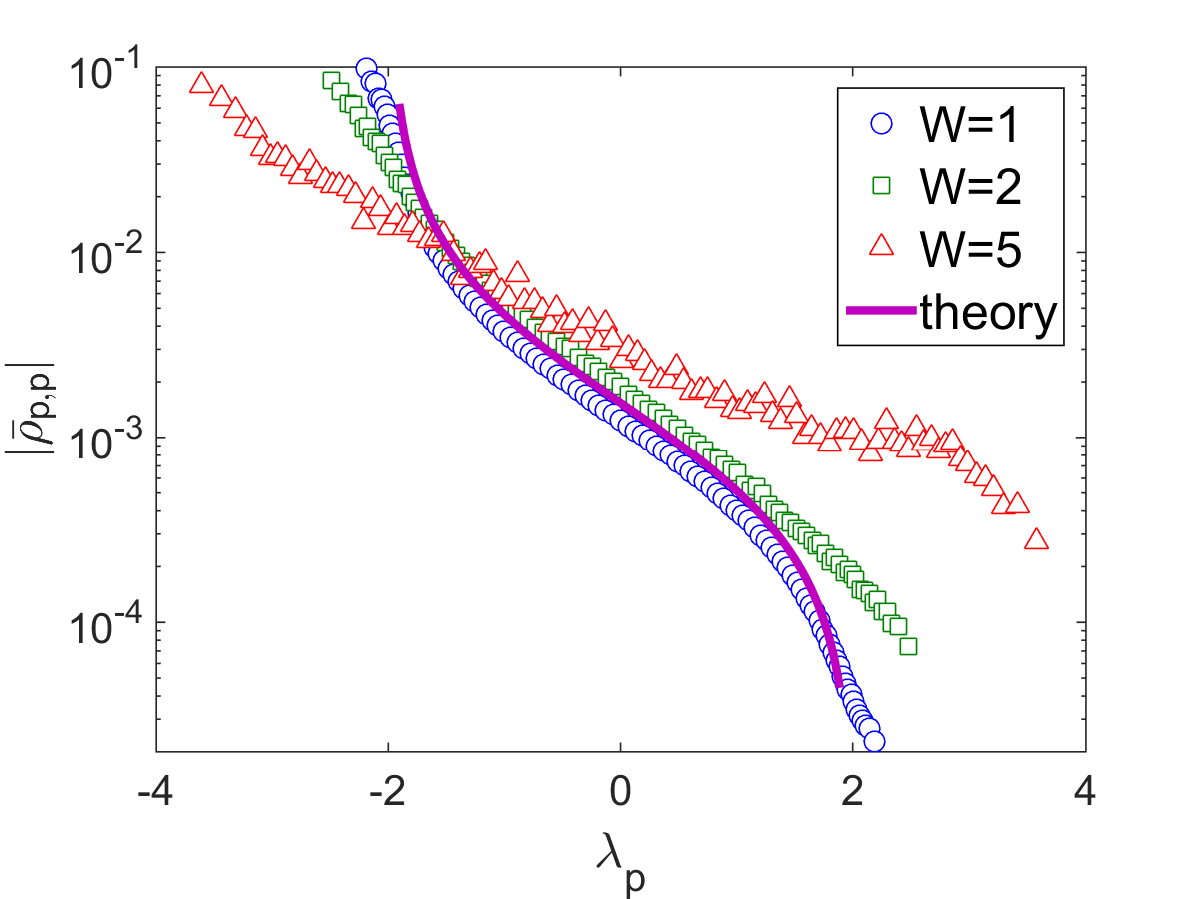}
\caption{Averaged absolute values at the diagonal of  $\varrho_{\infty}$, expressed
in the Anderson basis, as functions of averaged eigenvalues for the case of in-phase next-neighbor dissipation, $\alpha = 0$ and $l=1$ in Eq.~(\ref{eq:4}). Averaging was performed 
over $N_{\mathrm{r}}=10^2$ disorder realizations.  Theoretical prediction, Eq.~(\ref{eq:4e}), is shown by a thick solid line.
The other parameters are the same as in Fig.~1.} \label{fig:3a}
\end{figure}

We consider  an open single-particle model described by Eq.~(\ref{eq:1}) with a 
Hamiltonian \cite{anderson} 
\begin{align}
H=&\sum_k \epsilon_k b_k^{\dagger}b_k -(b_{k}^{\dagger} b_{k+1} + b_{k+1}^{\dagger} b_{k}), 
\label{eq:2}
\end{align}
where $\epsilon_k\in\left[-W/2, W/2\right]$ are random uncorrelated on-site energies, $W$ is the 
disorder strength, $b_k$ and $b_k^{\dagger}$ are the annihilation and creation operators of a boson on the $k$-th site.
We recall that the eigenvalues of the Hamiltonian are restricted to a finite interval, $\lambda_\nu \in \left[-2-W/2, 2+W/2 \right]$, while the 
respective eigenstates, $A_k^{(\nu)}$,  are exponentially localized.
The localization length is approximated by $\xi_\lambda \approx 24(4-\lambda^2)/W^2$  \cite{Thouless1979} with corrections about the band edges \cite{Derrida}.

%, while $n_k=b_k^{\dagger}b_k$ is the particle number operator on the same site. 
A single dissipative operator acts on a pair of sites,
\begin{equation}
V_k=(b_{k}^{\dagger} + e^{i \alpha}b_{k+l}^{\dagger})(b_{k}-e^{-i\alpha}b_{k+l}).
%~~~ l=\{1,2\} 
\label{eq:4}
\end{equation}
When $\alpha=0$, this operator tries  to synchronize the dynamics on the $k$ and $k+l$ sites, by constantly recycling
anti-symmetric out-of-phase mode into the symmetric in-phase one.
This  type of dissipation, with $l=1$, was introduced in Refs.~\cite{DiehlZoller2008,KrausZoller}.
A physical implementation of a Bose-Hubbard chain with neighboring sites coupled by such  dissipators
was discussed in Ref.~ \cite{marcos2012}. The proposed set-up consists of an array of superconductive resonators coupled by qubits;
a pair-wise dissipator  with $l=1$ and arbitrary phase $\alpha$ can be realized with the same set-up by \textit{(i)} varying the photonic mode frequency $\omega_k$ 
from cavity to cavity (to simulate the disorder), and \textit{(ii)} adjusting  position of the qubits
with respect to the centers of  the corresponding cavities \cite{adja}. In principle, this is enough for our purpose.
However, to address other possible implementations, we will consider also dissipators with $l > 1$.

%When {\color{blue}$\alpha=\pi/l$}, the operator works in the opposite way.

Next we assume $\varrho_{0}= \varrho_{N+1} = 0$ at the boundaries and analyze properties of the stationary solution $\varrho_{\infty}$ of Eq.~(\ref{eq:1}), 
which is unique due to the absence of relevant symmetries \cite{alicki,symmetry}. To find it,  
we use a column-wise vectorization of the  density matrix and define the asymptotic solution, a super-vector $\rho_{\infty}$,  as the kernel of
a Liouvillian-induced super-operator $\Pi$,  $\Pi\rho_{\infty} = 0$. 
After folding it into the matrix form  and trace-normalizing, we get the asymptotic state density matrix $\varrho_{\infty}$ \cite{exact}.

For the choice $\alpha = 0$ and $l=1$ in Eq.~(4), that is the case of in-phase next-neighbor dissipative coupling \cite{DiehlZoller2008,KrausZoller,marcos2012}, 
the  asymptotic density matrix 
exhibits a  patchy structure with several 'hot' localization spots, Fig.~\ref{fig:1}a.  
Remarkably, expressing the density operator in the basis of Anderson modes $A_{\nu}$, 
we get a near diagonal matrix with a strong contributions from the eigenstates from the lower part  of the  spectrum, Fig.~\ref{fig:1}b. 
%Moreover, the asymptotic state density matrix there, $\bar{\rho}$, is almost diagonal and exhibits localization, the diagonal elements decaying away from the lower band edge.
\begin{figure}
\includegraphics[width=\columnwidth]{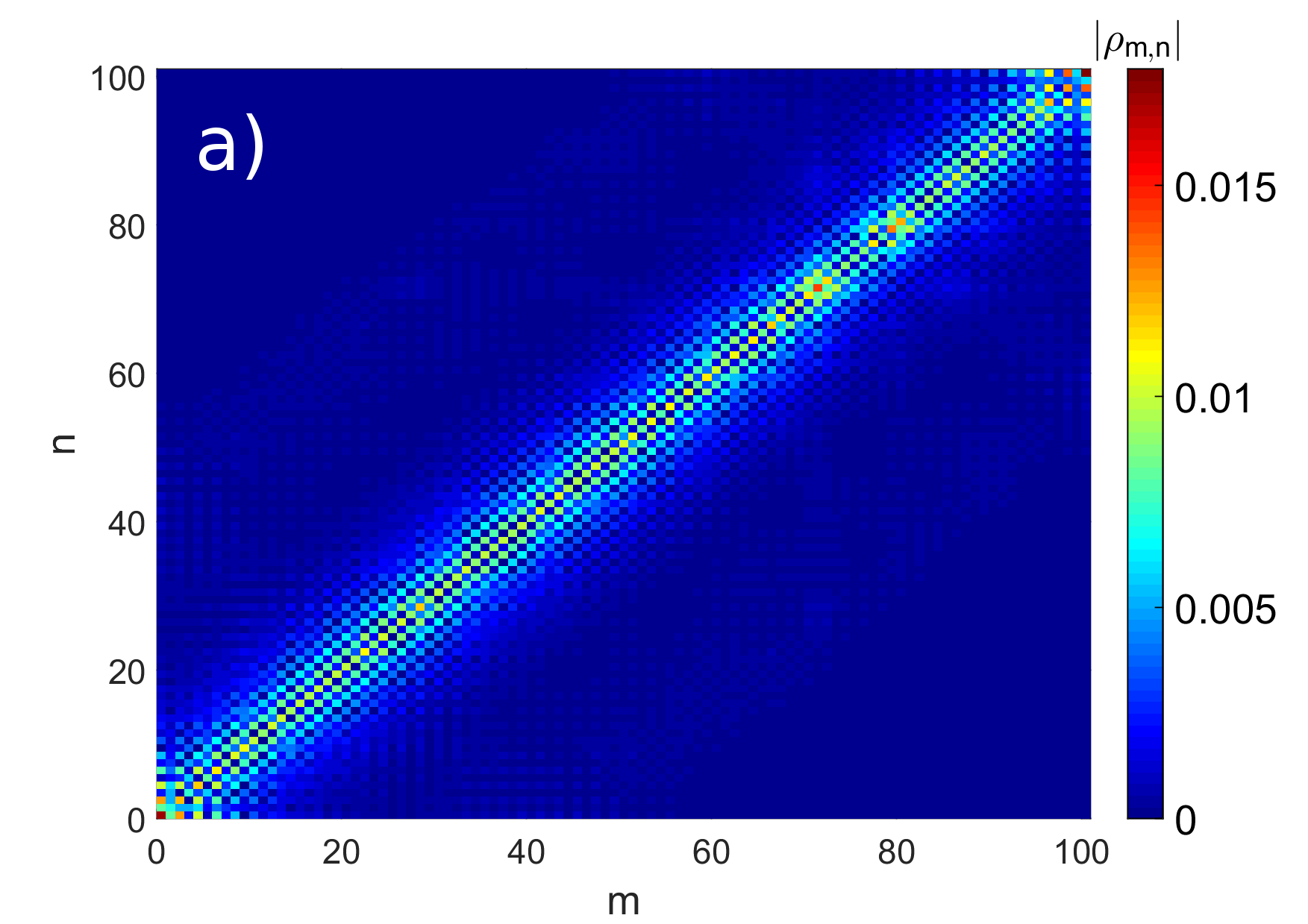}
\includegraphics[width=\columnwidth]{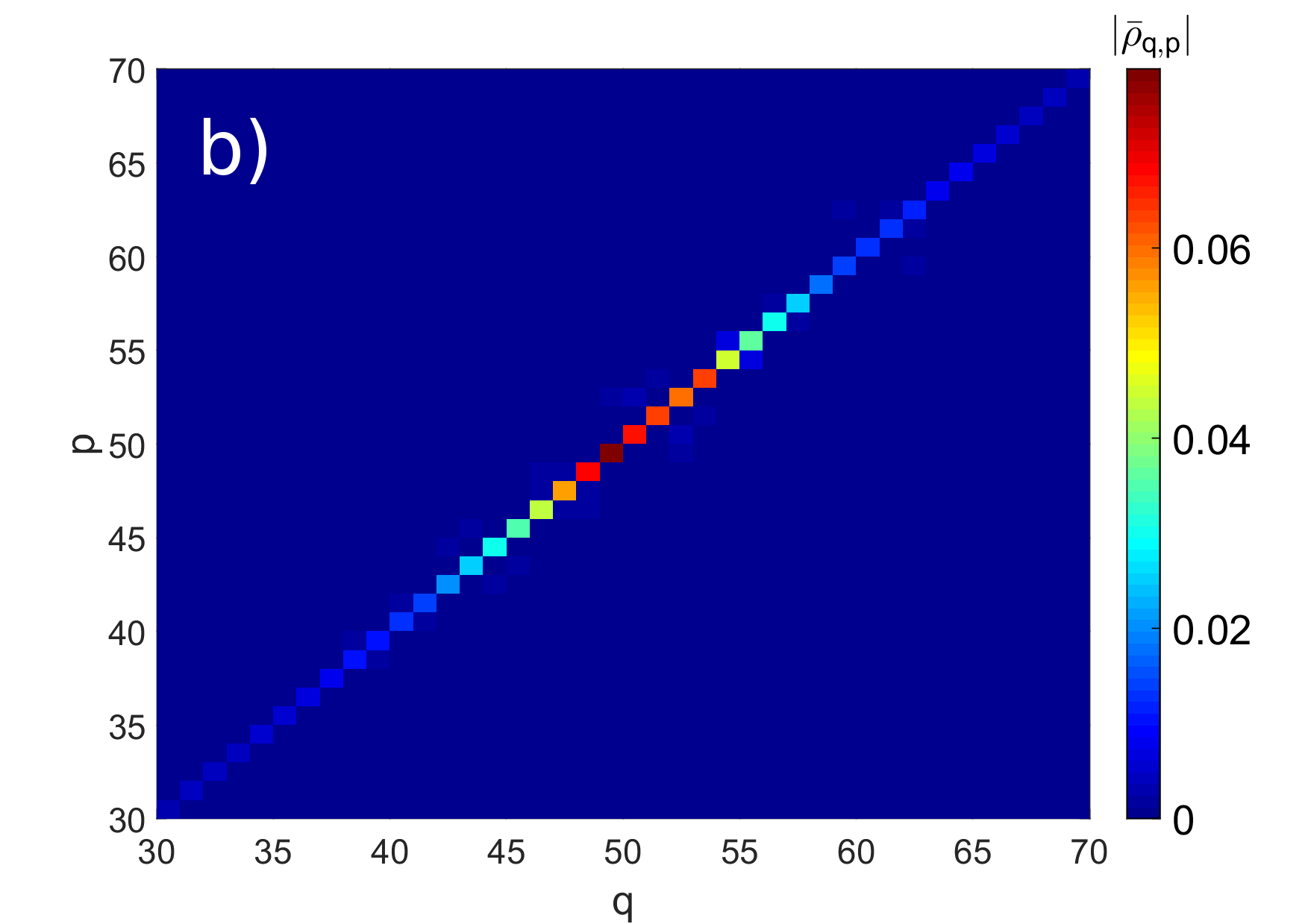}
\caption{(a) Absolute values of the elements of the  stationary density matrix  $\varrho_{\infty}$
in the direct basis and (b) the basis of the Anderson modes (a relevant part is shown) for a particular disorder realization of the model, Eqs.~(\ref{eq:1}-\ref{eq:2}),
and  out-of-phase next-nearest-neighbor dissipators, $\alpha = \pi$ and $l=2$ in Eq.~(4). Other parameters are  $W=1$, $N=100$, $\gamma =0.1$.} \label{fig:2}
\end{figure}
%********************************************
\begin{figure}
\includegraphics[angle=0,width=\columnwidth]{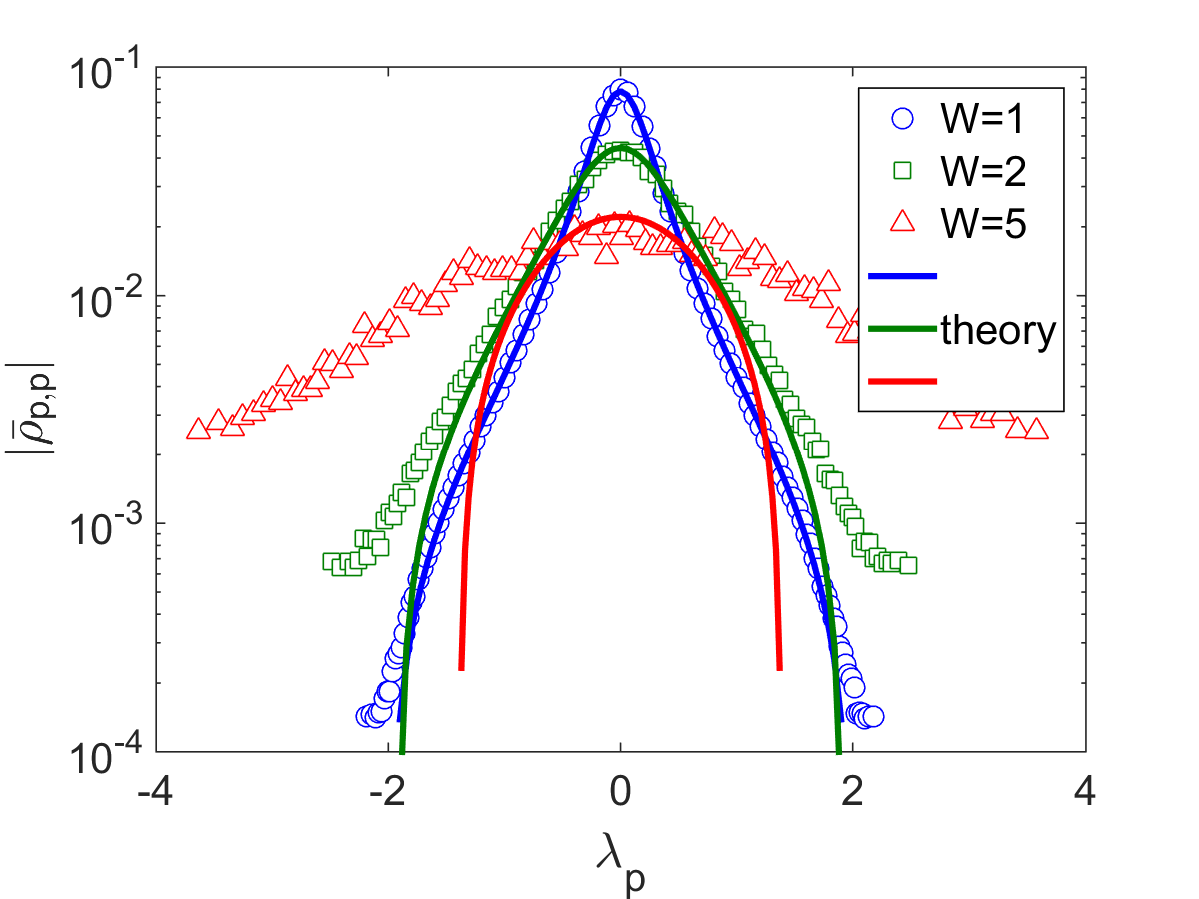}
\caption{Averaged absolute values of the diagonal elements
of $\varrho_{\infty}$, expressed in the Anderson basis, 
as functions of average eigenvalues for the case of  out-of-phase next-nearest-neighbor dissipator, $\alpha = \pi$ and $l=2$ in Eq.~(4). Averaging was performed
over $N_{\mathrm{r}}=10^2$ disorder realizations.  Theoretical predictions, Eq.~(\ref{eq:4e}), are shown by thick solid lines. Other parameters are 
$N=100$, $\gamma =0.1$.} \label{fig:3b}
\end{figure}
%********************************************************************
To evaluate this finding  analytically, 
we rewrite Eq.~(\ref{eq:1}) in the Anderson basis and neglect the off-diagonal elements.
%\cite{note1}. 
Under this approximation, the evolution of the diagonal elements is governed by dissipative terms only,
\begin{equation}
\label{eq:4a}
\dot{\bar{\varrho}}_{p,p}=\gamma\left(\sum_qI_{p,q}\bar{\varrho}_{q,q}-\bar{\varrho}_{p,p}\sum_q I_{q,p}\right),
\end{equation}
where the overlap coefficients $I_{p,q}=\sum_k|(\bar{V}_k)_{q,p}|^2$ are given by the dissipators in the Anderson basis, $\{\bar{V}_k\}$. Explicitly, 
\begin{equation}
\label{eq:4b}
I_{p,q}=\sum_k\left(A_{p,k+l}+e^{i\alpha}A_{p,k}\right)^2\left(A_{q,k+l}-e^{-i\alpha}A_{q,l}\right)^2.
\end{equation}
Denoting  $I_{p,k}^{\pm}=\left(A_{p,k+l}\pm e^{\pm i\alpha}A_{p,k}\right)^2$, we obtain  for the stationary solution: 
\begin{equation}
\label{eq:4c}
\bar{\varrho}_{p,p}=\frac{\sum_k I_{p,k}^{+} \sum_q I_{q,k}^{-} \bar{\varrho}_{q,q}}{\sum_k I_{p,k}^{-}\sum_q I_{q,k}^{+}}.
\end{equation}
The inner sums in the numerator and denominator do not depend on the index $p$, and subjected to the averaging over all eigenstates spanned by $q$. 
Because the disorder is spatially homogeneous,  the ensemble average 
makes the result also $k$-independent and so it corresponds to a  normalization constant.
With that we arrive at the following expression for the asymptotic density matrix:
\begin{equation}
\label{eq:4d}
\bar{\varrho}_{p,p}\propto\frac{\sum_k I_{p,k}^{+}}{\sum_k I_{p,k}^{-}}=\frac{I_{p}^{+}}{I_{p}^{-}},
\end{equation}
which is fully determined by the type of dissipation and a spatial structure of a particular  Anderson eigenstate.

For  in-phase next-neighbor dissipators, $\alpha=0$ and  $l=1$ in Eq.~(4), 
it follows $\sum_k\left(A_{p,k+1}\pm A_{p,k}\right)^2=2\pm2\sum_k A_{p,k+1}A_{p,k}=2\mp\lambda_p\mp\sum\epsilon_k A_{p,k}^2$ \cite{explain}. 
Except for the case of strong localization ($\xi_p\sim1$, for $W>4$ or about the band edges), the last term averages out due to  spatial disorder and can therefore 
be neglected, and so we end up with 
\begin{equation}
\label{eq:4e}
\bar{\varrho}_{p,p}\propto\frac{2-\lambda_p}{2+\lambda_p},
\end{equation}
This result explains the quick decay of the  contribution from the eigenstates away from the lower band edge. 
We numerically calculate, for different disorder strengths, the average distribution of the diagonal elements of $\varrho_{\infty}$,
expressed  in the Anderson basis, and plot them as functions of the average eigenvalues, see Fig.~\ref{fig:3a}. 
The obtained results are in a  good agreement with the theoretical prediction, Eq.~(\ref{eq:4e}). 
Note that the mismatch increases with the disorder strength $W$ and near the band edges; these effects follow from the nature of the made approximations.  

It is straightforward 
to see that in case of \textit{anti}-phase next-neighbor dissipators, $\alpha=\pi$ and  $l=1$, 
the symmetry of the problem leads to the inverse expression, 
$\bar{\varrho}_{p,p} \propto\frac{2+\lambda_p}{2-\lambda_p}$, 
and the asymptotic state localized near the upper band edge. 
A choice $\alpha=\pi/2$ makes all
dissipative operators $V_k$, Eq.~(\ref{eq:4}), Hermitian. This leads to the complete delocalization of 
the asymptotic state, $\varrho_{\infty}  = \id/N$. Intermediate phase values, $0<\alpha<\pi/2$ ($\pi/2<\alpha<\pi$), 
produce asympotic density matrix (expressed in the Anderson basis)  with dominating diagonal elements
localized near the lower (upper) band edge.

A qualitatively different  picture is observed with next nearest neighbor dissipators, $\alpha = \pi$  and  $l=2$. 
In this case the asymptotic state becomes delocalized in the original basis, Fig.~\ref{fig:2}a. 
At the same time, it remains localized in the Anderson basis, though shifted to the center of the spectrum, see Fig.~\ref{fig:2}b. 
Delocalization  in the direct space occurs due to the substantial spatial overlap of the contributing Anderson states, which are much weaker localized at the band center than at the edges. 
Analytically, that corresponds to $I_p^{-}=\sum_k\left(A_{p,k+2}+A_{p,k}\right)^2=\lambda_p^2-2\lambda_p\sum_k\epsilon_k A_{p,k}A_{p,k+1}+\sum_k \epsilon_k^2 A_{p,k}^2\approx\lambda_p^2+W^2/12$ 
and $I_p^{+}=4-I_p^{-}$,  and it follows  
\begin{equation}
\label{eq:4f}
\bar{\varrho}_{p,p}\propto\frac{4}{\lambda_p^2+W^2/12}-1.
\end{equation}
This expression indicates that the dominant contribution comes from the central Anderson modes. 
We also calculate the average profiles of asymptotic states 
for different disorder strengths to compare them with the analytical solution, 
which depends now explicitly on $W$, Fig.~\ref{fig:3b}. Again, we find a good agreement with the numerical results, while the mismatch increases 
with $W$ and distance from the band center. 

Is is noteworthy that in the limit of strong localization, $W \gg 1$, when all eigenstates are essentially single-site localized, dissipation induces strong delocalization. 
As it follows from Eq.~(\ref{eq:4b}), in this limit all overlap coefficients  become $I_{p,q} \sim \delta_{p,q}$, 
and the distribution of the values of the diagonal elements of the density matrix expressed in the Anderson basis should become near uniform. This 
means also delocalization in the direct space. As disorder strength increases, this trend can be seen on both Figs.~\ref{fig:3a} and \ref{fig:3b}.

%#########################################################################
%#                         Fig.2s:                        #
%#########################################################################
\begin{figure}
\begin{center}
\includegraphics[angle=0,width=\columnwidth,keepaspectratio,clip]{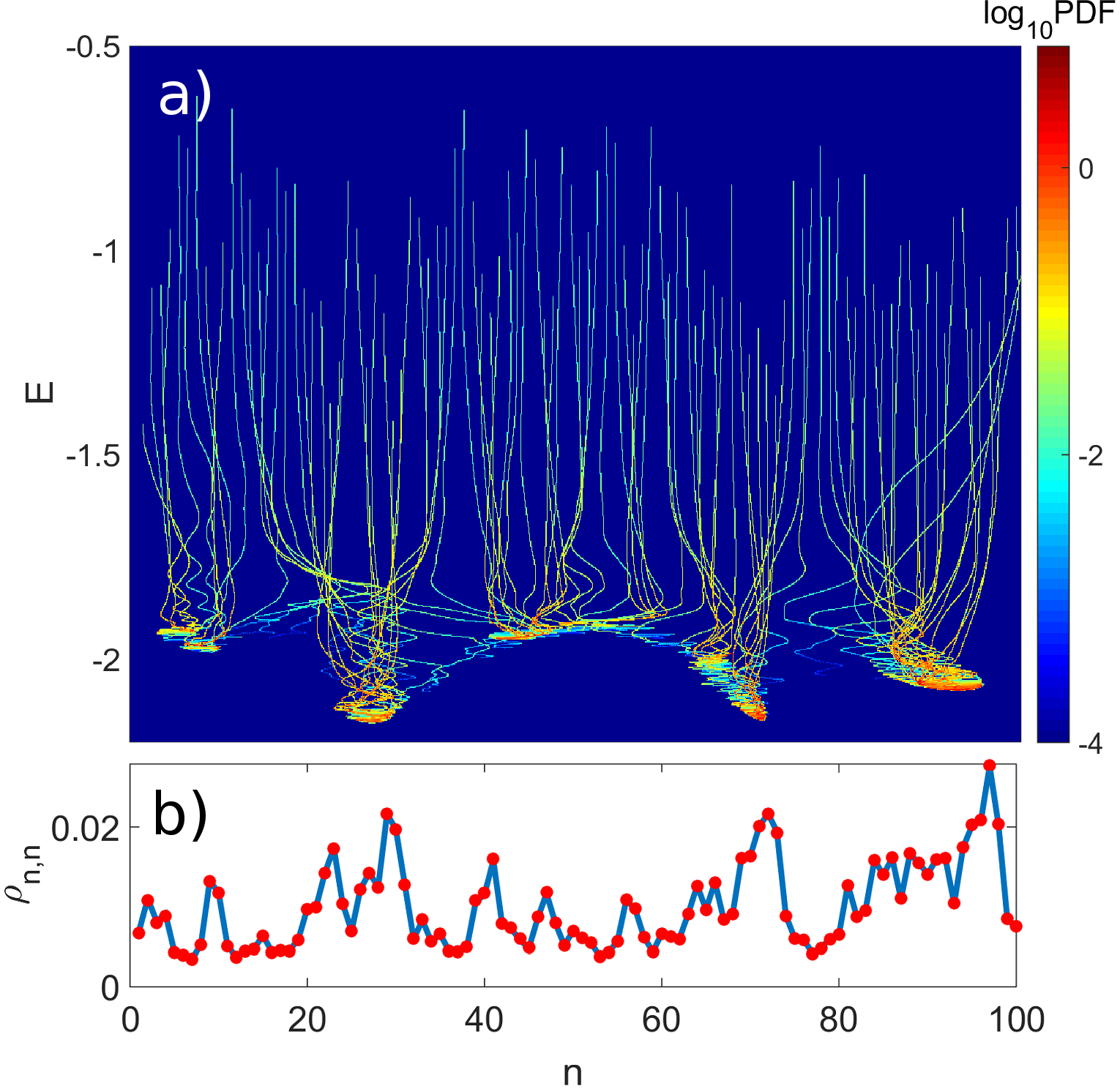}
\caption{(a) Probability density function on the position-energy plane
for quantum trajectories in the asymptotic regime; (b) The diagonal elements of the stationary
density matrix (expressed in the direct basis)  obtained as the 
stationary solution of Eq.~(\ref{eq:1}) (blue solid line) and as the marginal distribution of the above probability density function (red dots). 
The pdf was obtained with $M_{\mathrm{r}}=10^6$ trajectories and  transients time $t_{\mathrm{tr}}=10^4$. Other parameters are $W=1, \gamma=0.1, N=100, \alpha=0$, and $l=1$.}
\label{fig:4}
\end{center}
\end{figure}
%#########################################################################

We gain further insight in the dissipative effects by unraveling  deterministic  equation (1) 
into an ensemble of quantum trajectories \cite{plenio,dali,zoller1992}. 
This allows us to recast the evolution of the model system in terms of pure states,
governed by an effective non-Hermitian Hamiltonian, $\tilde{H} = H -\frac{i}{2}  \sum_k  V^\dagger_k V_k$, and random jumps induced by dissipators
$V_k$. This is not a formal step only; for some quantum optics realizations  it may properly model the reality of the experiment \cite{plenio}. 
For $W=1, \gamma=0.1$ and $N=100$, we choose  transient time $t_{\mathrm{tr}}=10^4$ and performed an 
averaging over $M_{\mathrm{r}}=10^6$ realizations to calculate
the probability density function (pdf)  on the position-energy plane, 
$\{n(t)=\sum_k\langle\psi(t)|n_k|\psi(t)\rangle$ and $E(t)=\langle\psi(t)|H|\psi(t)\rangle\}$. 

Figure \ref{fig:4}(a) presents the obtained pdf for the case of in-phase next-neighbor dissipators, $\alpha = 0$  and $l=1$.
On the trajectory level, the asymptotic dynamics of the system is remarkable. Several localized states are selected 
from the part of the spectrum specified by the phase properties of dissipators (here it is the lower band edge). The intermittent dynamics is 
a mixture of sticky-like beating near one of the localized eigenstates (non-Hermitian evolution with $\tilde{H}$ \cite{Fyodorov}),
which are interrupted by quantum jumps (induced by a randomly selected operator $V_k$).
Every jump  throws the system into the high-energy region from where it quickly relaxes,  through a fine-structured network, to one of the eigenstates. 
The structure of the network is specific to the disorder realization; however, it does not change with further increase of $M_{\mathrm{r}}$.
Marginal distribution over $n(t)$ recovers the diagonal elements of the asymptotic density matrix  $\varrho_{\infty}$ expressed in the direct basis,  see Fig.~\ref{fig:4}b.

We have shown that dissipation can
%play a constructive role in stabilizing Anderson localization and can 
be used to create steady states dominated by a few localized modes of a spatially disordered Hamiltonian.
Anderson modes are selected according to their spatial-phase properties inherited from the seeding plane waves, 
the eigenstates of the Hamiltonian in the zero-disorder limit \cite{ishii}, by using phase-parametrized dissipative operators.
It is possible to steer the system
into a desired  asymptotic state, with footprints of localization or completely delocalized, by changing  phase parameter of the dissipative operators.
%It is tempting to say that we observe 'localization over localization', i.e., an Anderson-like  localization in the basis of  Anderson modes.
%However, this localization is only of the power-law type, see Eqs.~(9-10). 

Our  findings pose several interesting issues for future investigations. First, it is a synergy between dissipation and modulation effects, such
as dynamical localization \cite{RefA1}. A quantum chaos, induced by strong periodic modulations \cite{RefA2,RefA3,RefA5,RefA6,RefA4} or quasi-periodic modulations \cite{RefA7}, 
can also play a role of an effective disorder and lead to AL-like effects. A controllable dissipation, added  to such systems,
can  lead to nontrivial results. Next direction is  many-body localization  \cite{Basko}, a phenomenon actively explored now.
What could be a result of the interplay of dissipation and many body localization  \cite{fish,les,les2} when the dissipative operators are phase-contolled? It is possible to create a MBL steady state
with local dissipators? To answer these questions, not only spectra of the MBL Hamiltonians and such integral characteristics as, e.g., 
the inverse participation ratio, have to be analyzed, but also spatial phase structure of MBL eigenstates.

{\it Acknowledgments:} This work was supported by the Russian Science Foundation grant No.\ 15-12-20029. 
%Computations we carried out on the Lobachevsky University supercomputer.


\begin{thebibliography}{1000}



% phase transitions
%\bibitem{sachdev} S. Sachdev, \textit{Quantum Phase Transitions}
%(Cambridge University Press, Cambridge, England, 1999).
\bibitem{Kramer1993} B. Kramer and A. MacKinnon, Rep. Prog. Phys. {\bf 56}, 1469 (1993).
\bibitem{Evers2008} F. Evers and A. Mirlin, Rev. Mod. Phys. {\bf 80}, 1355 (2008).
\bibitem{fifty} 50 Years of Anderson Localization, ed. by E. Abrahams (World Scientific, 2010).


\bibitem{fifty2} A. Lagendijk, B. van Tiggelen, and D. S. Wiersma, Physics Today \textbf{62}, 24 (2009).


\bibitem{Segev2013}  M. Segev, Y. Silberberg, and D. N. Christodoulides, Nature Photon. \textbf{7}, 197 (2013).
%\bibitem{Hu2008} H.~Hu, A.~Strybulevych, J.H.~Page, S.E.~Skipetrov, and B.A.~van~Tiggelen, Nature Phys. {\bf 4}, 945 (2008).
\bibitem{Billy2008} J.~Billy, V.~Josse, Z.~Zuo, A.~Bernard, B.~Hambrecht, P.~Lugan, D.~Cl\'{e}ment, L.~Sanchez-Palencia, P.~Bouyer, and A.~Aspect, Nature {\bf 453}, 891 (2008).
\bibitem{Roati2008} G.~Roati, C.~D'Errico, L.~Fallani, M.~Fattori, C.~Fort, M.~Zaccanti, G.~Modugno, M.~Modugno, and M.~Inguscio, Nature {\bf 453}, 895 (2008).
\bibitem{Kondov2011} S. S.~Kondov, W. R.~McGehee, J. J.~Zirbel, B.~DeMarco, Science {\bf 334}, 66 (2011).
\bibitem{Jen2012} F.~Jendrzejewski,	A.~Bernard,	K.~M\"{u}ller,	P.~Cheinet,	V.~Josse,	M.~Piraud,	L.~Pezz\'{e},	L.~Sanchez-Palencia,	A.~Aspect, and P.~Bouyer, Nature Phys. {\bf 8}, 398 (2012).
%\bibitem{nonlinear2} L. Sanchez-Palencia and M. Lewenstein, Nature {\bf 6}, 89 (2010).
%Deissler2010,Lucioni2011





\bibitem{book} H.-P. Breuer, F. Petruccione, \textit{The Theory of Open Quantum Systems} (Oxford University Press, Oxford, 2002).

%\bibitem{lucioni} B. Deissler {\it et al.}, Nat. Phys. {\bf 6}, 354 (2010); 
%E. Lucioni {\it et al.}, Phys. Rev. Lett. {\bf 106}, 230403 (2011).

% Dissipation & localization


% Dissipative localization

% - Dissipative quantum computing and dissipative engineering
% dissipative state engineering\rho\rho\rho\varrho
%\bibitem{wolf} F. Verstraete, M. M. Wolf, J. I. Cirac, Nature Phys. {\bf 5}, 633 (2009).

\bibitem{DiehlZoller2008} S. Diehl, A. Micheli, A. Kantian, B. Kraus, H. P. B\"uchler, P. Zoller, Nature Physics {\bf 4}, 878 (2008).
\bibitem{KrausZoller} B. Kraus H. P. B\"uchler, S. Diehl, A. Kantian, A. Micheli, P. Zoller, Phys. Rev. A {\bf 78}, 042307 (2008).

\bibitem{wolf2009} F.~Verstraete, M. M.~Wolf, and J.I.~Cirac, Nature Phys. {\bf 5}, 633 (2009).

%\bibitem{KlaersWeitz} J. Klaers, J. Schmitt, F. Vewinger, M. Weitz, Nature {\bf 468}, 545 (2010).
%% Referee B
%% Constructive effects of dissipation in bistable systems (e.g. stabilizing metastable states).
\bibitem{RefB1}  D.~Valenti,	L.~Magazz\'{u},	P.~Caldara,	and	B.~Spagnolo,	Phys. Rev. B {\bf 91}, 235412	(2015).
\bibitem{RefB2}  L.~Magazz\'{u},	D.~Valenti,	B.~Spagnolo, and M.~Grifoni,	Phys.	Rev. E	{\bf 92}, 032123	(2015).
\bibitem{RefB3}	L.~Magazz\'{u}, A.~Carollo,	B.~Spagnolo, D.~Valenti, J. of	Stat.	Mech. 054016 (2016).

% - localization (better say correlation of atoms, pairing in disorder-free lattice) due to dissipation
\bibitem{RefA8}  N.~Syassen, D.M.~Bauer, M.~Lettner, T.~Volz, D.~Dietze, J. J.~Garcia-Ripoll, J. I.~Cirac, G.~Rempe, S.~D\"{u}rr, Science, {\bf 320}, 1329 (2008).
% - dissipation and driving may increase purity and coherence of the condensate; constructive effect of dissipation; dimer
\bibitem{RefA10}  D.~Witthaut, F.~Trimborn, and S.~Wimberger Phys. Rev. Lett. 101, 200402 (2008).
% - dissipation and interaction induced localization; point-like dissipation; no disorder; B-H chain - discrete breathers, stabilization of BEC 
\bibitem{RefA11} D.~Witthaut, F.~Trimborn, H.~Hennig, G.~Kordas, T.~Geisel, and S.~Wimberger, Phys. Rev. A {\bf 83}, 063608 (2011).
\bibitem{RefA12}  G.~Kordas, S.~Wimberger, and D.~Witthaut, Phys. Rev. A {\bf 87}, 043618 (2013).




\bibitem{anderson} P. W. Anderson, Phys. Rev. \textbf{109}, 1492 (1958).
%\bibitem{Fyodorov} Y.V.~Fyodorov, JETP Letters {\bf 78}, 250 (2003).
%\bibitem{Huse} R.~Nandkishore, S.~Gopalakrishnan, and D.A.~Huse, Phys. Rev. B {\bf 90}, 064203 (2014).

%\bibitem{Frank2006} R.~Frank, A.~Lubatsch, J. Kroha, Phys. Rev. B \textbf{73},
 %245107 ({2006}).

%\bibitem{Yamilov2014} A. Yamilov, \emph{et~al.}, Phys. Rev. Lett. \textbf{112}, 023904 (2014).
%\bibitem{Basiri_2014} A. Basiri, Y. Bromberg, A. Yamilov, H. Cao, and T. Kottos,
%Phys. Rev. A \textbf{90}, 043815 (2014).


\bibitem{Stano2013} P.~Stano and P.~Jacquod, Nature Photonics \textbf{7}, 66 (2013).
\bibitem{LiuJ.2014} J.~Liu,	P. D.~Garcia,	S.~Ek,	N.~Gregersen,	T.~Suhr,	M.~Schubert,	J.~M\o rk,	S.~Stobbe, and P.~Lodahl, Nat. Nanotech. \textbf{9}, 285 (2014).
\bibitem{ivanchenko2015a} T. V.~Laptyeva, A. A.~Tikhomirov, O. I.~Kanakov and M. V.~Ivanchenko, Sci. Rep. {\bf 5}, 13263 (2015).
\bibitem{ivanchenko2015b} T. V. Laptyeva, S. V. Denisov, G. V. Osipov, M. V.  JETP Lett., {\bf 102} (9), 603 (2015).

%\bibitem{because} This is because the identity $\id$ commutes with
%all operators, including $H$ and all $V_k$, and therefore $\mathcal{L}_t(\id) = 0$.

\bibitem{les0} S. Genway, I. Lesanovsky, and J. P. Garrahan, Phys. Rev. E \textbf{89}, 042129 (2014)

\bibitem{alicki} R. Alicki, K. Lendi, 1987, \textit{Quantum Dynamical Semigroups and Applications},
Lecture Notes in Physics, Vol. 286 (Springer, Berlin).

\bibitem{fish} M. F. Fisher, M. Maksymenko, E. Altman, Phys. Rev. Lett. {\bf 116}, 160401 (2016).
\bibitem{les} E. Levi, M. Heyl, I. Lesanovsky, J. P. Garrahan, Phys. Rev. Lett. {\bf 116}, 237203 (2015).
\bibitem{les2} B. Everest, I. Lesanovsky, J. P. Garrahan, E. Levi, arXiv:1605.07019 (2016).


%Lindblad
%\bibitem{lind} G. Lindblad, Commun. Math. Phys. {\bf 48}, 119 (1976).
%\bibitem{gorini} V. Gorini, A. Kossakowski, E. C. G. Sudarshan, J. Math. Phys. {\bf 17}, 821 (1976).



% dissipative state engineering\rho\rho\rho\varrho
%\bibitem{wolf} F. Verstraete, M. M. Wolf, J. I. Cirac, Nature Phys. {\bf 5}, 633 (2009).

%\bibitem{DiehlZoller2008} S. Diehl, A. Micheli, A. Kantian, B. Kraus, H. P. B\"uchler, P. Zoller, Nature Physics {\bf 4}, 878 (2008).
%\bibitem{KrausZoller} B. Kraus H. P. B\"uchler, S. Diehl, A. Kantian, A. Micheli, P. Zoller, Phys. Rev. A {\bf 78}, 042307 (2008).
\bibitem{Bardyn} C. E. Bardyn, M. A. Baranov, C. V. Kraus, E. Rico, A. \.{I}mamo\'{g}lu, P. Zoller, and S. Diehl, New. J. Phys. {\bf 15}, 085001 (2013).
%\bibitem{KlaersWeitz} J. Klaers, J. Schmitt, F. Vewinger, M. Weitz, Nature {\bf 468}, 545 (2010).

%\bibitem{barr} J. T. Barreiro, M. M\"uller,	P. Schindler, D. Nigg, T. Monz,	M. Chwalla, M. Hennrich, C. F. Roos, P. Zoller, R. Blatt, Nature Phys. {\bf 6}, 943 (2010).
% this is the shortened version of the reference above:
\bibitem{barr} J. T. Barreiro, M. M\"{u}ller,	P. Schindler, \textit{et.\ al.}, Nature Phys. {\bf 6}, 943 (2010).


\bibitem{kienz} D. Kienzler, H.-Y. Lo, B. Keitch, L. de Clercq, F. Leupold, F. Lindenfelser, M. Marinelli, V. Negnevitsky, J. P. Home, Science {\bf 347} 53 (2015).
\bibitem{ketz} D. Vorberg, W. Wustmann, R. Ketzmerick, A. Eckardt, Phys. Rev. Lett. {\bf 111}, 240405 (2013).


\bibitem{marcos2012} D. Marcos, A. Tomadin, S. Diehl, and P. Rabl, New J. Phys. \textbf{15}, 055005 (2012).


\bibitem{adja} The relative position of a qubit in a cavity controls the phase of a complex coupling constant $q_k$
in the Jaynes-Cummings coupling term, $q_k^*b_k^{\dagger}\sigma_k^- + q_kb_k \sigma_k^+$, where qubit operator $\sigma_k^- = \arrowvert g_k\rangle\langle e_k|$ \cite{QO}.
To realize dissipator (4) with $l=1$ and $\alpha \neq 0$ in the set-up proposed in Ref.~\cite{marcos2012}, 
the coupling constant should vary with $k$ as $q_k = |q|\exp(-i\alpha k)$.

\bibitem{QO} P. Meystre and M. Sargent, \textit{Elements of Quantum Optics} (Springer, Berlin, 4 ed., 2007).
 

\bibitem{Thouless1979} D. J.~Thouless, In: \textit{Ill-condensed Matter}, Eds. R. Balian, R. Maynard, and G. Toulouse
(North-Holland, 1979).
\bibitem{Derrida} B. Derrida and E. Gardner, J. Physique {\bf 45}, 1283 (1984).


\bibitem{symmetry}  V. V.~Albert, L.~Jiang, Phys. Rev. A {\bf 89}, 022118 (2014).

\bibitem{exact}  Maximal absolute value of the elements in the r.h.s of equation (\ref{eq:1}) after substituion 
of $\varrho_{\infty}$ does not exceed $10^{-14}$.




%\bibitem{note1} According to the numerics, the approximation is violated in the case of strong dissipation, $\gamma \sim 1$, which is out the scope of the paper. 

% MBL
%\bibitem{Scardicchio} C.R. Laumann, A. Pal, and A. Scardicchio, Phys. Rev. Lett.
%{\bf 113}, 200405 (2014); C.L. Baldwin, C.R. Laumann, A. Pal, and A. Scardicchio, Phys. Rev. B {\bf 93}, 024202 (2016).

%\bibitem{kraus} A. Kossakowski, Rep. Math. Phys. {\bf 3}, 247 (1972).
%\bibitem{pre1997} S. Kohler, T. Dittrich, P. H\"anggi, Phys. Rev. E {\bf 55}, 300 (1997).


%\bibitem{spohn1} E.B. Davies, H. Spohn, J. of Stat. Phys. {\bf 19}, 523 (1978).
%\bibitem{lendi} K. Lendi, Phys. Rev. A {\bf 33}, 3358 (1986).

%\bibitem{katz} I. Katz, A. Retzker, R. Straub, R. Lifshitz, Phys. Rev. Lett.
%{\bf 99} 040404 (2007).
%\bibitem{prosen} T. Prosen, E. Ilievski, Phys. Rev. Lett. {\bf 107}, 060403 (2011).
%\bibitem{Shnirman} I. Kamleitner, A. Shnirman, Phys. Rev. B {\bf 84}, 235140 (2011).


%\bibitem{chan} C.-K. Chan, N. T. Lee, S. Gopalakrishnan, Phys. Rev. A {\bf 91}, 051601 (2015).

%\bibitem{mag} F. Haddadfarshi, J. Cui, F. Mintert, Phys. Rev. Lett. {\bf 114}, 130402 (2015).

%\bibitem{spohn} H. Spohn, Rep. Math. Phys. {\bf 10} 198 (1976).




%diehl dissipator
\bibitem{zoller} S. Diehl, A. Tomadin, A. Micheli, R. Fazio, P. Zoller, Phys. Rev. Lett. {\bf 105}, 015702 (2010).


%topologically driven
%\bibitem{Kitagawa} T. Kitagawa, E. Berg, M. Rudner, E. Demler, Phys. Rev. B {\bf 82}, 235114 (2010).


%\bibitem{ab1} P. Ponte, Z. Papi\'{c}, F. Huveneers, D. Abanin, Phys. Rev. Lett. {\bf 114}, 140401 (2015).

%\bibitem{laz} A. Lazarides, A. Das, R. Moessner, Phys. Rev. Lett. {\bf 115}, 030402 (2015).

%topologically driven

%\bibitem{PichlerZoller2010} H. Pichler, A. J. Daley, P. Zoller, Phys. Rev. A {\bf 82}, 063605 (2010).


% Quantum jump

\bibitem{explain}  
The last step relies on the identity obtained from the eigenstate equation $-(\lambda_p-\epsilon_k) A_{p,k}=A_{p,k-1}+A_{p,k+1}$, multiplied by $A_{p,k}$ and summed over $k$.

\bibitem{plenio}   M. B. Plenio and P. L. Knight, Rev. Mod. Phys. \textbf{70}, 101 (1998).

\bibitem{dali} J. Dalibard, Y. Castin, and K. M\o{}lmer, Phys. Rev. Lett. \textbf{68}, 580 (1992);

\bibitem{zoller1992}  R. Dum, A.S. Parkins, P. Zoller, and C.W. Gardiner, Phys. Rev. A \textbf{46}, 4382 (1992).

%\bibitem{purity} We have checked purity of several obatined asymptotic states; it did not exceed $0.35$.

\bibitem{Fyodorov} Y. V.~Fyodorov, JETP Letters {\bf 78}, 250 (2003).

               
\bibitem{ishii} K. Ishii, Prog. Theor. Phys. Suppl. \textbf{53}, 77 (1973).

\bibitem{RefA1}  S.~Fishman, D. R.~Grempel, and R. E.~Prange, Phys. Rev. Lett. {\bf 49}, 509 (1982).
\bibitem{RefA2}  S.~Wimberger, A.~Krug, and A.~Buchleitner, Phys. Rev. Lett. {\bf 89}, 263601 (2002).
\bibitem{RefA3}  F.~J\"{o}rder, K.~Zimmermann, A.~Rodriguez, and A.~Buchleitner, Phys. Rev. Lett. {\bf 113}, 063004 (2014).



\bibitem{RefA5} J. E.~Bayfield, G.~Casati, I.~Guarneri, and D.W.~Sokol, Phys. Rev. Lett. {\bf 63}, 364 (1989).
\bibitem{RefA6} C. F.~Bharucha, J.C.~Robinson, F. L.~Moore, B.~Sundaram, Q.~Niu, and M.G.~Raizen, Phys. Rev. E {\bf 60}, 3881 (1999).
\bibitem{RefA4} E. J.~Galvez, B. E.~Sauer, L.~Moorman, P. M.~Koch, and D.~Richards, Phys. Rev. Lett. 61, 2011 (1988).
\bibitem{RefA7} J.~Chab\'{e}, G.~Lemari\'{e}, B.~Gr\'{e}maud, D.~Delande, P.~Szriftgiser, and J. C.~Garreau, Phys. Rev. Lett. {\bf 101}, 255702 (2008).

\bibitem{Basko} D. M. Basko, I. L. Aleiner and B. L. Altshuler, Ann. Phys. {\bf 321}, 1126 (2006).

% Dynamical localization


% MBL

%\bibitem{Anderson80} L. Fleishman and P. W. Anderson, Phys. Rev. B {\bf 21}, 2366 (1980).

%\bibitem{Altshuler1997} B.L.~Altshuler, Y.~Gefen, A.~Kamenev, and L.S.~Levitov, Phys. Rev. Lett. 78, 2803 (1997).

%\bibitem{Basko} D. M. Basko, I. L. Aleiner and B. L. Altshuler, Ann. Phys. {\bf 321}, 1126 (2006).

%\bibitem{gorn} I.V.~Gornyi, A.D.~Mirlin, and D.G.~Polyakov, Phys. Rev. Lett. {\bf 95}, 206603 (2005). 

%\bibitem{huse0} V.~Oganesyan and D.A.~Huse, Phys. Rev. B {\bf 75}, 155111 (2007).

%\bibitem{Altshuler2010} I.~Aleiner, B.~Altshuler, and G. V.~Shlyapnikov, Nature Physics {\bf 6} (11), 900 (2010).

%\bibitem{huse}  J. Choi, S. Hild, J. Zeiher, P. Schau\ss{}, A. Rubio-Abadal, T. Yefsah, V. Khemani, D. A. Huse, I. Bloch, C. Gross, Science  \textbf{352}, 1547 (2016).



%R. Nandkishore and D. A. Huse,  Annual Review of Condensed Matter Physics \textbf{6}, 15 (2015).

%\bibitem{huse1}  J. Smith, A. Lee, P. Richerme,	B. Neyenhuis, P. W. Hess, P. Hauke, M. Heyl, D. A. Huse, and C. Monroe, Nature Physics  \textbf{12},  907 (2016).






\end{thebibliography}
\end{document}